\documentstyle[prl,aps,floats]{revtex}
\input{psfig.tex}

\tighten
\begin{document} 
\draft
\twocolumn[\hsize\textwidth\columnwidth\hsize\csname@twocolumnfalse\endcsname
\preprint{Imperial/TP/96-97/51}
\title{Early structure formation with cold plus hot dark matter 
--- a success of strings plus inflation models}
\author{Richard A. Battye{$^{1}$}, Jo\~{a}o Magueijo{$^{2}$} 
and Jochen Weller{$^{2,3}$}}
\address{
${}^1$ Department of Applied Mathematics and Theoretical Physics, University of
Cambridge, \\ Silver Street, Cambridge CB3 9EW, U.K. \\ 
${}^2$ Theoretical Physics Group, Blackett Laboratory, Imperial
College, Prince Consort Road, London SW7 2BZ, U.K. \\
${}^3$ Department of Physics, University of California, Davis, CA
95616, U.S.A.}
\maketitle
\begin{abstract}
Quantum fluctuations created during inflation can
account for the observed matter distribution in the linear regime if
the universe has two components of dark matter, one which is cold and
collisionless, and the other which is hot and free streams on small
scales. However, this free streaming property of the hot component prevents
early structure formation, and since objects, such as damped
Lyman-$\alpha$ systems, have been observed at
high redshift, it is necessary to produce more power
on small scales. Here, we show that the situation can be improved substantially in models where cosmic strings are formed at the end of inflation,
and in which both inflation and strings participate in the 
generation of structure.
\end{abstract}

\date{\today}

]

\renewcommand{\thefootnote}{\arabic{footnote}}
\setcounter{footnote}{0}

During the next decade various high precision measurements will
enhance our understanding of the universe immeasurably. Two satellites
(MAP and PLANCK) will map the cosmic microwave background (CMB) over a
wide range of scales and ground-based redshift surveys (for example,
Sloan Digital Sky Survey and 2Df) should measure the clustering of
luminous matter. Using these complementary measurements --- as well as many others (for example, those of Type Ia supernovae and light element abundances) --- it is hoped that robust estimates can be made for a large number of cosmological parameters. However, underlying this ambitious enterprise is the origin of the primordial fluctuations, which is usually assumed to be a near scale invariant spectrum of adiabatic density perturbations created by quantum effects during some de-Sitter phase associated with cosmic inflation.

Under various assumptions as to the matter content of the universe
three broad classes of models based on inflation have been suggested
to fit the current observations of galaxy clustering in the linear
regime ($\lambda>30h^{-1}{\rm Mpc}$) when normalized to the
amplitude of anisotropies in the CMB detected by the COsmic
Background Explorer (COBE) satellite (see, for example,
ref.~\cite{GawSilk}). Each of these models is based primarily on a
universe whose matter density is dominated by cold dark matter (CDM)
particles ($\Omega_{\rm m}\sim\Omega_{\rm c}$), such as
neutralinos or axions, with a much
 smaller component of baryons as predicted by Big Bang Nucleosynthesis
$\Omega_{\rm b}\approx 0.05-0.1$. At  present, possibly the most
popular of these classes of models have a matter density which is less
than critical ($\Omega_{m}\approx 0.3-0.5$),  with in one class
($\Lambda$CDM) the shortfall from critical being made up by a non-zero
cosmological constant $(\Omega_{\Lambda}+\Omega_{\rm m}=1$) and the
other (OCDM) having an open topology. The final class of models (CHDM)
~\cite{CHDM} have $\Omega_{\rm m}=1$ with an extra relativistic or
hot dark matter (HDM) component ($\Omega_{\nu}\approx 0.2-0.3$), such
as massive neutrinos, which, since they free-stream, reduce the amount
of power on intermediate and small scales, allowing the model to fit the shape of the observed power spectrum in the linear regime. Reconciling the predictions of this cold plus hot dark matter model with observations at high redshift is the subject of this {\it letter}.

The relevant observations concerning us are those of damped Lyman-$\alpha$ (DLYA) systems~\cite{DLYA} which are thought to probe structure
as it was for $z>2$ in the linear regime, and on scales smaller 
than otherwise possible. There seems to be a conflict between 
the measured neutral gas fraction, particularly at very high redshifts
($z\approx 4$), and the predictions made for CHDM
models~\cite{MB,KBHP}, which --- due to the free-streaming property
of the HDM component --- tend to lack of power on such small
scales. This is one of  the main reasons why, in spite
of the superior achievements of CHDM models on linear scales, OCDM and 
$\Lambda$CDM models are now preferred. 

For many years the most popular alternative to the standard adiabatic
scenarios discussed above has been to form structure by accretion of
matter onto topological defects (for a review, see
ref.~\cite{review}). After much recent
work~\cite{turok,us,shellard,mag}, it appears that, at least when
$\Omega_{\rm m}=1$, these models appear to suffer from the converse
problem; they have a tendency to produce insufficient power on large
scales to match the current observations when normalized to COBE, but
are known to produce copious amounts of small scale
power\cite{us,shellard,mag}, even in scenarios with a HDM
component \cite{note}. 
This is particularly true in the case of
cosmic strings and one may wonder if the complementary failings of CHDM based on inflation and string models could be erased in a mixed string plus inflation CHDM model. 

The possibility that cosmic structure could be due to both strings and inflation~\cite{rachel} is a very real one which has been the topic of much recent research~\cite{mag1,stinf,stinf1}, since it could arise 
in a number of well motivated inflationary scenarios~\cite{linde,dterm,lr,CLLSW,yoko}. The details of this paper are only weakly sensitive to the specifics of the inflationary model and then only in an essentially predictable way. However, to give an example, let us consider 
D-term inflation~\cite{dterm} (see ref.~\cite{LyRio98} for a review). 
These models make use of supersymmetry in order to provide 
cosmology with a flat potential, as required by inflation, without 
the need for a finely tuned coupling constant. During inflation, the
inflaton moves along this flat direction, which is a direct result of
the U(1) symmetries in the model, and  at the end of inflation the underlying symmetry becomes broken, producing cosmic strings via the Kibble mechanism during the subsequent phase transition.

In any string plus inflation scenario the large angle CMB
anisotropies, which are normalized to COBE, are due both to strings
and inflation with a model dependent ratio between the two; it  being
a non-trivial function of the number of $e$-foldings during inflation,
the inflationary energy scale and that for the symmetry breaking phase
transition which produced the strings. Given a specific model it would
be possible to relate this to parameters of the model as was done in
refs.\cite{mag1,stinf1}. In both cases it was shown that any value of
this ratio was possible for sensible model parameters and therefore
for the purposes of this paper we will leave it arbitrary. In
particular we will use the notation of ref.~\cite{stinf1}, where the
power spectra were added as, for example, $P(k)=\alpha P^{\rm
inf}(k)+(1-\alpha)P^{\rm str}(k)$, where $0\le\alpha\le 1$ and $P^{\rm
inf}(k)$, $P^{\rm str}(k)$ are individually computed spectra for the
inflation and string models normalized to COBE. The specific
spectra that we shall use have been computed using
CMBFAST~\cite{cmbfast} and in the string case the model for the defect
stress energy used in refs.~\cite{us,stinf1}. This has been shown to
represent many of the features of a realistic string network, although
it may under estimate the amount of small-scale power
produced~\cite{shellard,mag}. Any extra small-scale power will further
improve the situation relative the observations being discussed here.

The neutral gas fraction in DLYA systems has been computed using a
variety of techniques for the standard CHDM scenario based on
inflation. Initially~\cite{MB,KBHP} the Press-Schechter (PS)
approximation~\cite{PS} was used to make an estimate and more
recently~\cite{LLSSV,MBHWK,GKWH,HSR} hydrodynamical simulations have
improved upon this, although the basic picture has remained the same. In attempting to estimate the size of the effect
of including a string component one is faced with a number of
obstacles to making a quantitative prediction, mainly related to the
non-Gaussian nature of the perturbations involved. This makes creating
a realization of initial conditions for a hydrodynamic simulation
almost impossible without an accurate simulation of the defect network
and of course one of the basic premises of the PS
formalism is that the fluctuations initially form a Gaussian random field. It
has been suggested~\cite{pdf} that a simple generalization of the PS
approximation can be made by replacing the Gaussian probability
distribution function (PDF) with that computed from a simulation and,
indeed, it has been shown that such a modified theory yields good
predictions for particular PDFs using $N$-body
simulations~\cite{james}. A PDF has been computed for the pure string
component~\cite{james,ASWA} showing that it has positive skewness. 
In the mixed scenarios considered here, the total PDF is the
convolution of the individual components, and so the total cumulants are
the sum of the individual cumulants. Hence, the combined PDF
will also have positive skewness.

Here, we will repeat
the standard PS calculation assuming Gaussian initial conditions
following closely the calculation of ref.~\cite{EH}. Since the
measured skewness for the cosmic strings PDF is positive, one might
expect that this calculation would act as  lower bound on the gas
fraction created in a given scenario. However, as we shall see the
effects of non-Gaussianity are subtle and this is not always the
case. To illustrate this we will use a simple non-Gaussian
distribution with a PDF which is a log-normal distribution, as
suggested in ref.~\cite{james}.

DLYA systems are observed as wide absorption troughs in the spectra of
distant quasars. These troughs indicate that the line of sight to the
quasar intersects a region of neutral hydrogen (HI) with a column
density $\ge 10^{20} {\rm cm}^{-2}$. For high redshift objects
($z\ge2$) there are hints, that these absorption lines are produced in
turbulent protospheroids \cite{Lanzetta:95}, which are the natural
progenitors of galaxies. The
measured abundance of damped Lyman-$\alpha$ systems at redshift
$z=4$ \cite{obs}, implies that  the neutral gas
density $\Omega_{\rm gas}$ in units of the critical density is
$\Omega_{\rm gas}\left(z\approx 4\right)h = \left(9.3 \pm
3.8\right) \times 10^{-4}$, in a universe which has $\Omega_{\rm m}=1$, 
where the Hubble constant is parametrized in the usual way
$H_0=100\,h\,{\rm km}\,{\rm sec}^{-1}\,{\rm Mpc}^{-1}$.

The neutral gas fraction in DLYA
systems is given by $\Omega_{\rm gas}= f_{\rm HI}\Omega_{\rm b}{\rm
erfc}\left({\delta_c}/\sqrt{2}\sigma_{R}\right)$, which is the result of the integration over the tail of the underlying
Gaussian PDF. In this expression
$0<f_{\rm HI}\le 1$ is the fraction of hydrogen which is neutral, $\Omega_{\rm b}$ is
the average  baryon density relative to critical, $\delta_{\rm c}(\approx 1.69)$ is the
density contrast required for collapse to take place and $\sigma_{R}$
is the rms density fluctuation in a ball with size of the DLYA systems.
To estimate this scale
we assume that the systems are  protospheriods which are about to collapse to
rotationally supported gaseous discs. In this  spherical collapse
scenario~\cite{Peebles} the comoving size of the system is related to the
circular velocity, $V_{\rm c}$, by 
\begin{equation}
R=86h^{-1}{\rm kpc}\left(\frac{V_{\rm c}}{50{\rm km}\,{\rm sec}^{-1}}\right)\left(\frac{5}{1+z}\right)^{1/2}\,.
\end{equation}
A typical velocity is $V_c\sim 50 \,{\rm km}\, {\rm sec}^{-1}$,
which corresponds  to the minimal mass ($\sim 10^{10}h^{-1}M_\odot$)
needed for spherical collapse to take place~\cite{H}. The rms
fluctuations on this scale, $\sigma_R$, are computed in the standard
way by integration over a window function corresponding to a spherical
top-hat of size $R$.

\begin{figure}[h]
\setlength{\unitlength}{1cm}
\centerline{\vbox{
\psfig{figure=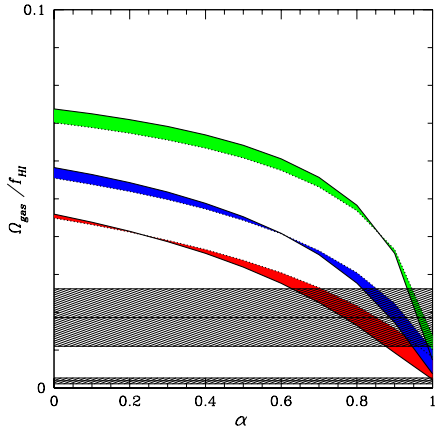,height=6cm,width=6cm}
}}
\caption{The neutral gas density $\Omega_{\rm gas}/f_{\rm HI}$ 
at $z=4$ plotted as a function of $\alpha$ for a CHDM model with
$\Omega_\nu=0.2$.  The three curved shaded regions are bounded by the
Gaussian PDF (solid line) and a non-Gaussian PDF with $A=0.5$ (dotted
line) and correspond to the region in which the actual value for the
mixed scenario actually lies. The circular velocities of the three
regions are given by $V_c=75{\rm km\, sec}^{-1}$ (top), $50{\rm km\, sec}^{-1}
$ (middle) and $25{\rm km\, sec}^{-1}$ (bottom). 
The parallel shaded regions correspond to the observational limits 
if $f_{\rm HI}=1.0$ (bottom) and $f_{\rm HI}=0.1$ (top). 
Recall that $\alpha=1$ corresponds to pure adiabatic perturbations and
$\alpha=0$ to strings only. Note that the non-Gaussian
PDF gives a lower value, albeit only slightly, than the
Gaussian case for certain values of $\alpha$. This non-intuitive effect is explained in the text.
}
\label{fig-omega2}
\end{figure}

In fig.~\ref{fig-omega2} we have plotted $\Omega_{\rm gas}/f_{\rm HI}$
at $z=4$ computed assuming Gaussian fluctuations (solid line) as a
function of $\alpha$ for models with $\Omega_{\rm c}=0.7$, $\Omega_{\rm
b}=0.1$, $\Omega_\nu=0.2$ and $h=0.5$ --- those parameters which were
suggested to give the best fit to the observed galaxy clusterings on large
scales in ref.~\cite{GawSilk}. Simulations suggest that realistically one
might expect $f_{\rm HI}\approx 0.1$~\cite{MBHWK,GKWH}, but if
one were wanting to be 
conservative in ruling out a scenario,  $f_{\rm HI}=1$ can be used to
compute an absolute upper bound on $\Omega_{\rm gas}$. If $\alpha=1$,
that is, just adiabatic fluctuations, then one sees that for all the values of
$V_c$ used, the upper bounds are compatible with the
observations; the
larger values of $V_c$ producing more neutral hydrogen than lower
ones. However, if $f_{\rm HI}=0.1$ none of these are compatible with the
value of $\Omega_{\rm gas}$ detected and this can be seen to be true for a
wide range of cosmological parameters~\cite{EH}. Even more striking from
this figure is that the observed value can be achieved very easily by the
inclusion of only a very small string component $\alpha\approx 0.8$, and
in fact the predictions could be made compatible with even smaller values of
$f_{\rm HI}$.    

A simple non-Gaussian distribution which has been suggested to
represent the PDF in cosmic string models~\cite{james} is the
log-normal distribution given by 
\begin{equation}
p_A(y)={C\over\sqrt{2\pi A^2}}\exp\left[-{1\over 2}x^2(y)-Ax(y)\right]\,,
\end{equation}
for $y>-C/B$ and zero otherwise, where $Ax(y)=\log(Cy+B)$,
$B=\exp(A^2/2)$ and $C=\sqrt{B^4-B^2}$. This has a single parameter
$A(>0)$, with the limiting case $A=0$ corresponding to a Gaussian
distribution. In this case, one can deduce that the neutral gas
fraction is given by 
\begin{equation}
\Omega_{\rm gas}={f_{\rm HI}\Omega_{\rm b}{\rm erfc}\left[\displaystyle{1\over
A\sqrt{2}}\log\left(C{\delta_{\rm c}\over
\sigma_{R}}+B\right)\right]\bigg{/}{\rm erfc}\left[\displaystyle{A\over2\sqrt{2}}\right]}\,.
\end{equation}
It was suggested in ref.~\cite{james} that $A\approx 0.17$ on cluster
scales $(R\sim 10h^{-1}{\rm Mpc}$), but on the smaller scales under
consideration here the value is likely to be somewhat larger~\cite{ASWA}, say
$A\approx 0.5$. We have also included computations of $\Omega_{\rm
gas}$ for a model using this PDF and $A=0.5$ in fig.~\ref{fig-omega2}, to
illustrate the effects of non-Gaussianity. Intuitively, one might have
expected the effects of introducing non-Gaussianity to increase
$\Omega_{\rm gas}$, but it appears this is not always the case. Using a
non-Gaussian PDF does enhance the creation of more rare objects, that
is $\sigma_{R}>>\delta_c$. However, if $\sigma_{R}\sim \delta_c$, as
can be the case when $\alpha$ is small , then the overall
normalization, which allows the integration over the whole PDF account
for all the matter in the universe, is reduced.

\begin{figure}[h]
\setlength{\unitlength}{1cm}
\centerline{\vbox{
\psfig{figure=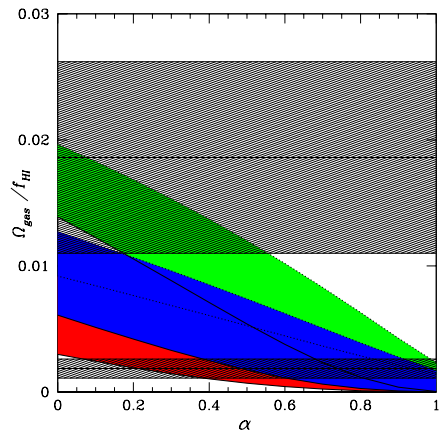,height=6cm,width=6cm}
}}
\caption{The same quantities as plotted in  fig.~\ref{fig-omega2} 
but for $\Omega_\nu=0.3$. Note that none of the models are even
compatible with observations of $\Omega_{\rm gas}$ for
$\alpha=1$, and that the inclusion of the  effects of
non-Gaussianity is now to increase $\Omega_{\rm gas}$ in contrast to the case of $\Omega_{\nu}=0.2$.} 
\label{fig-omega3}
\end{figure}

Fig.~\ref{fig-omega3} contains the equivalent information for the
larger value of $\Omega_{\nu}=0.3$ (and now $\Omega_{\rm
c}=0.6$). Clearly, such values of $\Omega_{\nu}$ are incompatible with
the observations if $\alpha=0$ since even the upper bound on
$\Omega_{\rm gas}$ is woefully short of the observed gas fraction. However, for
string-inflation admixtures of around $50\%$ each the upper bound on
the gas fraction becomes within range of the observations if one
assumes Gaussian fluctuations, albeit with $f_{\rm HI}>0.1$. The
effects of non-Gaussianity are now more intuitive, since
$\sigma_R<<\delta_c$ and the DLYA systems can be thought of as being
rare objects once again.

Therefore, we have shown that CHDM model can be made consistent with high
redshift observations, if we are prepared to resort to well motivated
strings plus inflation scenarios. The physical origin of this result
is the extra fluctuations on small scales, due to the strings, but
also non-Gaussianity can play an important role. If the DLYA systems
can be thought of as rare objects ($\delta_c>>\sigma_R$), then this can
enhance their production relative to Gaussian. But if they are formed
as $1\sigma$ fluctuations then their production is slightly
suppressed in non-Gaussian models. It is clear that the effects of
non-Gaussianity will be more prevalent at early times when $\sigma_R$
is small  and that less objects will formed later as $\sigma_R$
increases. This has interesting implications for the epoch of galaxy
and cluster assembly~\cite{BMWb}.

To conclude we have suggested that the CHDM scenario, and  more importantly
$\Omega_m=1$, can be resuscitated by the inclusion of a string
component since this allows structure to be formed earlier than in the
pure adiabatic case. Apart from the flaw of the CHDM
models which we have discussed here, there are other
observations which favour the $\Lambda$CDM model and
OCDM models mainly because they have $\Omega_{\rm m}<1$. It would be
interesting to understand how these observations would be modified for
the non-Gaussian theories discussed here. It is clear that
dynamical measures, such as the evolution of X-ray cluster abundance,
could all be made compatible with $\Omega_{\rm m}=1$ is a sufficiently
non-Gaussian model. Work on the observational status of this class of models continues.

We would like to thank P. Ferreira, J. Robinson and A. Liddle for
helpful comments. The computations were done at the UK National
Cosmology Supercomputing Center supported by PPARC, HEFCE and Silicon
Graphics/Cray Research. We acknowledge financial support from Trinity
College (RAB) the Royal Society (JM) and the German Academic Exchange Service (DAAD) (JW).

\def\jnl#1#2#3#4#5#6{\hang{#1, {\it #4\/} {\bf #5}, #6 (#2).} }
\def\jnltwo#1#2#3#4#5#6#7#8{\hang{#1, {\it #4\/} {\bf #5}, #6; {\it
ibid} {\bf #7} #8 (#2).} } 
\def\prep#1#2#3#4{\hang{#1, #4.} } 
\def\proc#1#2#3#4#5#6{{#1 [#2], in {\it #4\/}, #5, eds.\ (#6).} }
\def\book#1#2#3#4{\hang{#1, {\it #3\/} (#4, #2).} }
\def\jnlerr#1#2#3#4#5#6#7#8{\hang{#1 [#2], {\it #4\/} {\bf #5}, #6.
{Erratum:} {\it #4\/} {\bf #7}, #8.} }
\def\prl{Phys.\ Rev.\ Lett.}
\def\pr{Phys.\ Rev.}
\def\pl{Phys.\ Lett.}
\def\np{Nucl.\ Phys.}
\def\prp{Phys.\ Rep.}
\def\rmp{Rev.\ Mod.\ Phys.}
\def\cmp{Comm.\ Math.\ Phys.}
\def\mpl{Mod.\ Phys.\ Lett.}
\def\apj{Ap.\ J.} 
\def\apjl{Ap.\ J.\ Lett.}
\def\aap{Astron.\ Ap.}
\def\cqg{Class.\ Quant.\ Grav.} 
\def\grg{Gen.\ Rel.\ Grav.}
\def\mn{MNRAS}
\def\ptp{Prog.\ Theor.\ Phys.}
\def\jetp{Sov.\ Phys.\ JETP}
\def\jetpl{JETP Lett.}
\def\jmp{J.\ Math.\ Phys.}
\def\zpc{Z.\ Phys.\ C}
\def\cupress{Cambridge University Press}
\def\pup{Princeton University Press}
\def\wss{World Scientific, Singapore}
\def\oup{Oxford University Press}

\pagebreak
\pagestyle{empty}


\begin{thebibliography}{99}

\bibitem{GawSilk}
\jnl{E. Gawiser and J. Silk}{1998}{}{Science}{280}{1405}

\bibitem{CHDM}
\jnl{A. Klypin, J. Holtzman, J. Primack and E. Regos}{1993}{}{\apj}{416}{1}

\bibitem{DLYA}
\jnl{A.M. Wolfe, D.A Turnshek, H.E. Smith and R.D. Cohen}{1986}{}{\apj~
Supplements}{61}{249}

\bibitem{MB}
\jnl{C.P. Ma and E. Bertschinger}{1994}{}{\apj}{434}{L5}

\bibitem{KBHP}
\jnl{A. Klypin, S. Borgani, J. Holtzman and J. Primack}{1995}{}{\apj}{444}{1}

\bibitem{review}
\book{A. Vilenkin \& E.P.S. Shellard}{1994}{Cosmic strings and other topological defects}{\cupress} \jnl{M. Hindmarsh \& T.W.B. Kibble}{1995}{}{\it Rep. Prog. Phys.}{58}{477}

\bibitem{turok}
\jnl{U.L. Pen, U. Seljak and N. Turok}{1997}{}{\prl}{79}{1615}

\bibitem{us}
\jnl{A. Albrecht, R.A. Battye and J. Robinson}{1997}{}{\prl}{79}{4736}
\jnl{R.A. Battye, J. Robinson and  A. Albrecht}{1998}{}{\prl}{80}{4847}
\jnl{A. Albrecht, R.A. Battye and J. Robinson}{1999}{}{\pr}{D59}{023508}

\bibitem{shellard}
\jnl{B. Allen, R.R. Caldwell, S. Dodelson, L. Knox, E.P.S. Shellard
and A. Stebbins}{1997}{}{\prl}{79}{2624}\jnl{P.P. Avelino,
E.P.S. Shellard, J.H.P. Wu and B. Allen}{1998}{}{\prl}{81}{2008}\prep{J.H.P. Wu, P.P. Avelino, E.P.S. Shellard and B. Allen}{1998}{}{astro-ph/9812156}

\bibitem{mag}
\jnl{C. Contaldi, M. Hindmarsh and J. Magueijo}{1999}{}{\prl}{82}{679}
\bibitem{note}
We should note that this is not a robust feature
of every active model for structure formation, but it appears to be
true for the most studied examples: the cosmic string and texture scenarios.

\bibitem{rachel} 
\jnl{R. Jeannerot}{1997}{}{\pr}{D56}{6205}

\bibitem{mag1}
\jnl{C. Contaldi, M. Hindmarsh and J. Magueijo}{1999}{}{\prl}{82}{2034}

\bibitem{stinf}
\prep{P.P. Avelino, R.R. Caldwell, and C.J.A.P. Martins}{1998}{}
{astro-ph/9809130}

\bibitem{stinf1}
\prep{R.A. Battye and J. Weller}{1998}{}
{astro-ph/9810203}

\bibitem{linde}
\jnl{A. Linde}{1991}{}{\pl}{259B}{38}
\jnl{A. Linde}{1994}{}{\pr}{D49}{748}

\bibitem{dterm}
\jnl{J.A. Casas, J.M. Moreno, C. Mu\~noz, and M. Quir\'os}{1989}{}{\np}{B328}{272}\jnl{E. Halyo}{1996}{}{\pl}{387B}{43}
\jnl{P. Bin\'etruy and G. Dvali}{1996}{}{\pl}{388B}{241}

\bibitem{lr}
\jnl{A. Linde and A. Riotto}{1997}{}{\pr}{D56}{1841}

\bibitem{CLLSW}
\jnl{E.J. Copeland, A.R. Liddle, D.H. Lyth, E.D. Stewart and
D. Wands}{1994}{}{\pr}{D49}{6410}

\bibitem{yoko}
\jnl{J. Yokoyama}{1988}{}{\pl}{212B}{273}\jnl{J. Yokoyama}{1989}{}{\prl}{63}{712}

\bibitem{LyRio98}
\prep{D.H. Lyth and A. Riotto}{1998}{}{hep-ph/9807278}

\bibitem{cmbfast}
\jnl{U. Seljak and M. Zaldarriaga}{1996}{}{\apj}{469}{437}

\bibitem{PS}
\jnl{W.H. Press and P. Schechter}{1974}{}{\apj}{187}{452}

\bibitem{LLSSV}
\jnl{A.R. Liddle, D.H. Lyth, R.K. Schaefer, Q. Shafi and P.T.P. Vianna}{1996}{}{\mn}{281}{531}

\bibitem{MBHWK}
\jnl{C.P. Ma, E. Bertschinger, L. Hernquist, D.H. Weinberg and N. Katz}{1997}{}{\apj}{484}{L1}

\bibitem{GKWH}
\jnl{J.P. Gardner, N.Katz, D.H. Weinberg and L. Hernquist}{1997}{}{\apj}{486}{42}

\bibitem{HSR}
\prep{M.G. Haehnelt, M. Steinmetz and M. Rauch}{1997}{}{astro-ph/9706201}

\bibitem{pdf}
\jnl{W.A. Chiu, J.P. Ostriker and M.A. Strauss}{1998}{}{\apj}{494}{475}

\bibitem{james}
\prep{J. Robinson and J. Baker}{1999}{}{astro-ph/9905098}

\bibitem{ASWA}
\prep{P.P. Avelino, E.P.S. Shellard, J.H.P. Wu and B. Allen}{1998}{}{astro-ph/9803120}

\bibitem{EH}
\prep{D.J. Eisenstein and W. Hu}{1997}{}{astro-ph/9710252}

\bibitem{Lanzetta:95}
\jnl{K.M. Lanzetta, A.M. Wolfe and D.A. Turnshek}{1995}{}{\apj}{440}{435}

\bibitem{obs}
\jnl{L.J. Storrie-Lombardi, R.G. McMahon and
M.J. Irwin}{1996}{}{\mn}{283}{L79}
\jnl{L.J. Storrie-Lombardi, M.J. Irwin and
R.G. McMahon}{1996}{}{\mn}{282}{1330}

\bibitem{Peebles}
\book{P.J.E. Peebles}{1980}{The Large Scale Structure of the
Universe}{Princeton University Press}

\bibitem{H}
\jnl{M.G. Haehnelt}{1995}{}{\mn}{273}{249}

\bibitem{BMWb}
\prep{R.A. Battye, J. Magueijo and J. Weller}{1999}{}{In progress}

\bibitem{BF}
\prep{N.A. Bahcall and X. Fan}{1998}{}{astro-ph/9803277}

\end{thebibliography}
\end{document}